\documentclass[aps,twocolumn]{revtex4}
\usepackage{graphicx}
\usepackage{gensymb}
\usepackage{amsmath}

\begin{document}
\def\bb{\begin{equation}}
\def\ee{\end{equation}}

\title{Attosecond Transient Absorption Spectroscopy of Strongly Correlated Mott Insulators: Signature of the Creation and Annihilation of Double Occupancy}

\author{Youngjae Kim$^{*}$}

\affiliation{
School of Physics, KIAS, Seoul 02455, Korea
}

\date{\today}

\begin{abstract}
We applied the time-resolved attosecond transient absorption spectroscopy to systematically investigate ultrafast optical responses of condensed matter systems. 
Under an intense pump pulse, absorption spectra indicate that the non-interacting electrons of band insulators produce a field-induced redshift, known as the dynamical Franz-Keldysh effect, as commonly expected. 
In contrast to the band insulators, in Mott insulators, unconventional spectra are observed which do not fully reflect the dynamical Franz-Keldysh effect. 
While it still exhibits the fishbone-like structures mimicking the dynamical Franz-Keldysh effect, the spectra show a negative difference absorption below the band edge, rendering a blueshift.
In addition, the decomposed difference absorption reveals the creation and the annihilation of double occupancy mainly contribute to the negative signal, implying that the unconventional spectra are purely driven by the electron-correlations. 
These demonstrations of unconventional responses would guide us to the correlation-embedded attosecond electron dynamics in condensed matter systems.
\end{abstract}

\maketitle

Since ultrashort laser technologies have grown over the past decades it has reached the attosecond (as) temporal resolution (1 as $=$ $10^{-18}$ s), which allows access to new insights in the time domain of 10 $-$ 100 as shorter than the typical time scale of electron dynamics\cite{Kienberger1,Cavalieri1,Krausz1,Lucchini0}. 
As a result, attosecond transient absorption spectroscopy (ATAS) has been enabled to naturally capture the electron dynamics in matters and has become an essential technique to investigate highly nonlinear responses and their transient spectral properties at extreme time scales within sub-cycle laser pulses\cite{Holler1,Wang1,Goulielmakis1}. 
In its early stage, ATAS has been widely used to demonstrate the dynamics of atomistic systems\cite{Wang1,Goulielmakis1,Michael1,Drescher1}, and now it has been extended to condensed matter systems, to explorer crystal symmetries\cite{Sato1}, electronic structures renormalizations\cite{Schultze1}, dipole responses\cite{Mashiko1}, measuring photocarriers\cite{Schlaepfer1,Zurch1}, and the dynamical Franz-Keldysh effect (DFKE)\cite{Yacoby,Jauho,Novelli1,Otobe1,Du1,Lucchini1,Lucchini2}. 

Among these studies, one of the essential achievements would be the control of fundamental properties to establish electronic responses directly by using an external electric field.
When a condensed matter is exposed to an external intense electric field, the resulting absorption spectra are modified because of field-induced additional channels leading to a redshift below the band edge, referred to as the Franz-Keldysh effect (FKE)\cite{Franz,Keldysh,Tharmalingam,Nahory}.

Recent studies on a temporally extended scheme of the FKE under an alternating electric field, known as the DFKE, has been evolved with pump-induced intraband transitions to give an intriguing feature of real time phase oscillations, the fishbone structures, in addition to the redshift of the FKE\cite{Lucchini2,Sato1}. It also has been observed that the dynamical nature of the DFKE can be characterized in the crossover regime of the adiabatic parameter, $\gamma_{a} (= U_{pon}/\omega_{p}) \sim 1$, the center of competition between the pump photon energy ($\omega_{p}$) and the pondoromotive energy ($U_{pon}$).
Then the FKE begins to appear in the adiabatic regime, i.e., $\gamma_{a} \gg 1$\cite{Novelli1,Otobe1,Gordon1}. 
Several studies on the DFKE in condensed matters have been reported from the experiments on a diamond where the absorption modified by the intense pump pulse was directly captured by probe (XUV)\cite{Lucchini1}. The results were also supported by the theoretical approach of time-dependent density functional theory (TDDFT) to show the oscillating property of absorption as well as the phase delay. 

More recently, nontrivial natures found in the attosecond transient reflectivity have been deeply resolved to incorporate the excitonic dynamics being an atomistic attribute into condensed matter systems\cite{Lucchini0}, which implies that the combined natures could have a significant role to provide unexpected insight of attosecond dynamics. 
In addition, a study of the ATAS of multi-band charge transfer insulators within the framework of TDDFT+U pointed out that screening would affect the DFKE to be changed\cite{Dejean1}. 
However, the responses of ATAS combined with pure Mott phase still remains unclear. 
Considering the preceding results, it would be timely to suggest an exquisite signature from the ATAS if the dynamics of condensed matter systems are purely associated with rather strong electron-correlation.

In this study, by employing the real-time exact diagonalization method, we shall demonstrate the ATAS to explorer unbiased results of attosecond dynamics from band insulators and Mott insulators. 
In the band insulators, an intense pump pulse induces a conventional modification of absorption corresponding to the DFKE, which trivially accompanies the redshift below the band edge and the fishbone structures as a result of the intraband electron motion. 
In contrast to the DFKE, however, in the Mott insulator where the insulating band structures purely originated from the electron-correlation, it is observed from the ATAS that unconventional modification in their spectra appears to show a blueshift below the band edge violating the DFKE, but the fishbone-like structures still exist to mimic the DFKE. 
Furthermore, we investigate the decomposed difference spectra of the ATAS from the Mott insulators, and reveal that the transition pathways associated with creation and annihilation of double occupancy, the special state due to on-site repulsions, evidently contributes to the blueshift. 
Understanding of the unconventional ATAS for the Mott insulator provides insight on dynamical linking between the electron-correlation and the attosecond responses, which guides us to the correlation-embedded attosecond dynamics in condensed matters.

We write model Hamiltonians\cite{Maeshima} to describe the one-dimensional periodic solids with finite size $L = 8$, which reads,
\begin{eqnarray}
{\cal H}_B &=&-t_{h}\sum_{i,\sigma}(c^{\dagger}_{i,\sigma}c_{i+1,\sigma}+h.c.) + \Delta /2 \sum_{i,\sigma} (-1)^{i}
\end{eqnarray}
for band insulators where the fully non-interacting electrons are considered and $\Delta$ corresponds to the staggered crystal potential. And,
\begin{eqnarray}
{\cal H}_M &=&-t_{h}\sum_{i,\sigma}(c^{\dagger}_{i,\sigma}c_{i+1,\sigma}+h.c.) + U \sum_{i,\sigma} n_{i,\sigma}n_{i,-\sigma}
\end{eqnarray}
for Mott insulators where the electrons are strongly correlated via the Hubbard $U$ for the on-site repulsion.
The $c^{\dagger}_{i\sigma}$ ($c_{i\sigma}$) is the creation (annihilation) operator for an electron with spin ${\sigma}$ at site $i$, the hopping $t_{h}$, and the number operator $n_{i,\sigma} = c^{\dagger}_{i,\sigma}c_{i,\sigma}$.

After the half-filled ground state $|\Psi(\tau\rightarrow-\infty)\rangle=|\Psi_{0}\rangle$ is obtained, we perform the time-dependent calculation under a given light-wave by solving the time-dependent Schr\"{o}dinger equation, $i\partial/\partial\tau|\Psi(\tau)\rangle = {\cal H}_{B(M)}(\tau)|\Psi(\tau)\rangle$, within the 4th-order Runge-Kutta algorithm\cite{Lee1}. 
The time-dependence in the Hamiltonian is evolved from the Peierls phase\cite{Kaneko}, $t_{h}$ $\rightarrow$ $t_{h}(\tau) = t_{h}e^{-iA(\tau)/c}$, with a vector potential $A(\tau)$ and the speed of light $c$. 
To evaluate the absorption spectra, we carry out a real time current density\cite{Kaneko,Sato2} $J(\tau) = \langle \Psi(\tau)|j(\tau)|\Psi(\tau)\rangle$ with a current operator $j(\tau) = -c \partial / \partial A(\tau) {\cal H}_{B(M)}(\tau)$. Then, one can obtain the real part of the absorption as,
\begin{eqnarray}
\sigma_{1}(\omega) = {Re}[J(\omega)/E(\omega)]
\end{eqnarray}
from the current $J(\omega) = \int d\tau e^{-\eta\tau+i\omega\tau} J(\tau)$ and electric field $E(\omega) = \int d\tau e^{-\eta\tau+i\omega\tau} E(\tau)$ with the relation of $E(\tau) = -(1/c) \partial / \partial \tau A(\tau)$ and the smear parameter $\eta / t_{h} = 0.4$. The probe is defined as $E(\tau)=E_{pr}\delta(\tau)$ with $E_{pr} \ll 1$ to guarantee a linear response.

\begin{figure}
\includegraphics[width=0.41\textwidth]{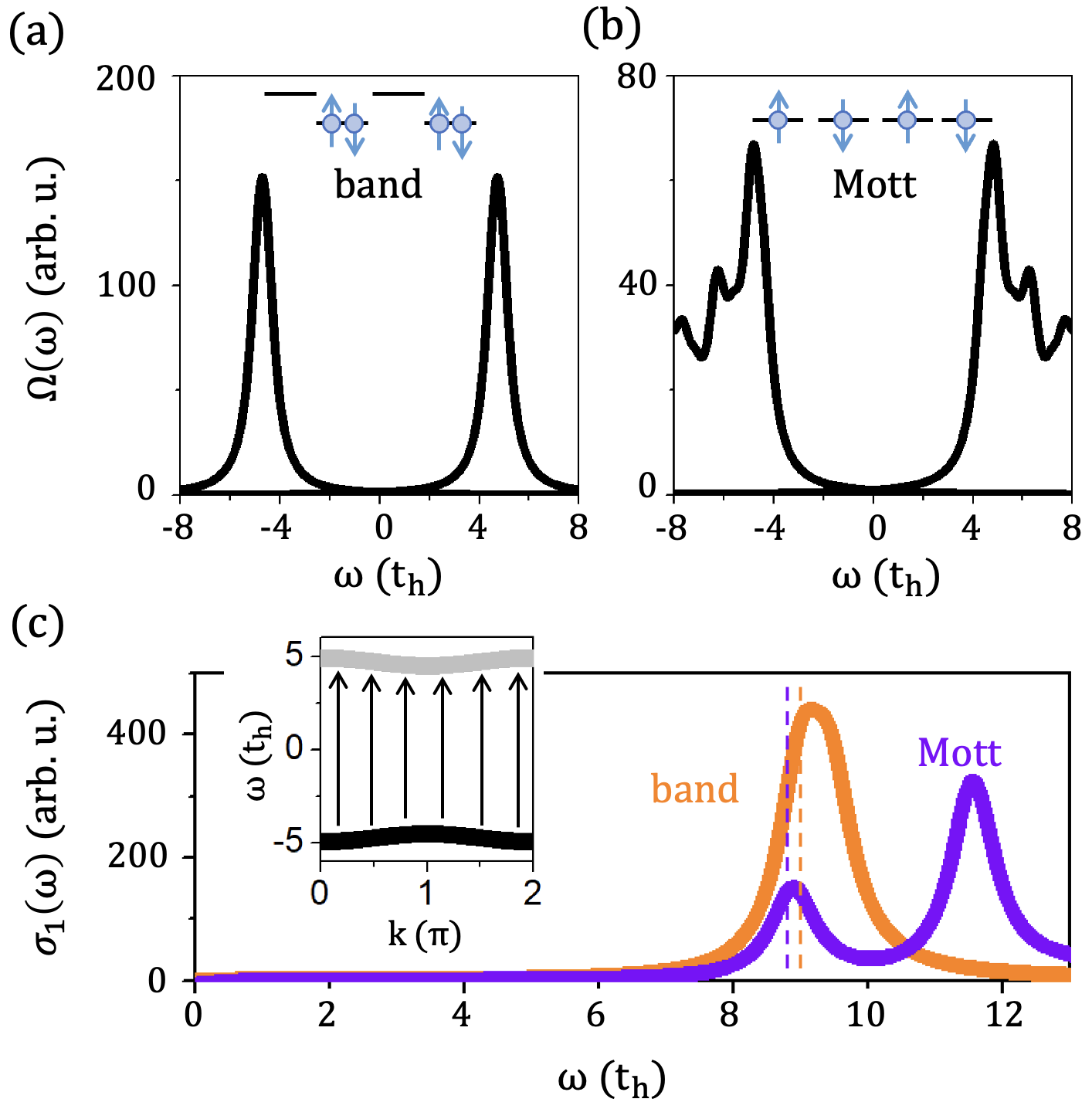}
\caption{
Calculated spectral weights and absorption spectra. 
(a),(b), Spectral weight of band insulators (a) and Mott insulators (b). 
The schematics in (a),(b) denote the ground electronic configurations and site energies.
(c), The ground state absorption spectra $\sigma_{1}(\omega)$ of the band insulators and the Mott insulators.
We fix the parameters of the staggered potential, $\Delta/t_{h} = 9$, for (a) and the Hubbard potential, $U/t_{h} = 12$, for (b), respectively, hereafter. The inset in the (c) depicts the probe-induced absorption process from the valence to the conduction bands. The dashed lines indicate the frequencies of each band edge.
}
\label{FIG1}
\end{figure}

In the Fig.\ref{FIG1}, we display the ground state properties of the band insulator and the Mott insulator. 
Under given parameters $\Delta$ and $U$, the energy scales represented by the spectral weights have a similar insulating phase to each other as seen from the Fig.\ref{FIG1}(a),(b). 
The spectral weights are extracted by the Lehmann representation\cite{Gora},
$\Omega(\omega) = Im|(1/\pi)\sum_{i,\sigma} [\langle \Psi_{0}|c^{\dagger}_{i,\sigma}{1/(\omega+E_{0}-{\cal H}_{B(M)}+i\eta)}c_{i,\sigma}|\Psi_{0}\rangle-\langle\Psi_{0}|c_{i,\sigma}{1/(\omega+E_{0}-{\cal H}_{B(M)}+i\eta)}c^{\dagger}_{i,\sigma}|\Psi_{0}\rangle]$, and the $E_{0}$ denotes the ground state eigenvalue corresponding to the ground eigenstate $|\Psi_{0}\rangle$. 
Strong correlations in the Mott insulator lead to slightly wider bandwidths than that of the band insulator.
In the Fig.\ref{FIG1}(c), the absorptions are found to have the lowest transitions at similar frequencies of band edges, $\omega/t_{h}=9$ for the band insulator and $\omega/t_{h}=8.8$ for the Mott insulator, respectively.
Hence the energy scales for ATAS on both systems would be similar to each other.

\begin{figure}
\includegraphics[width=0.47\textwidth]{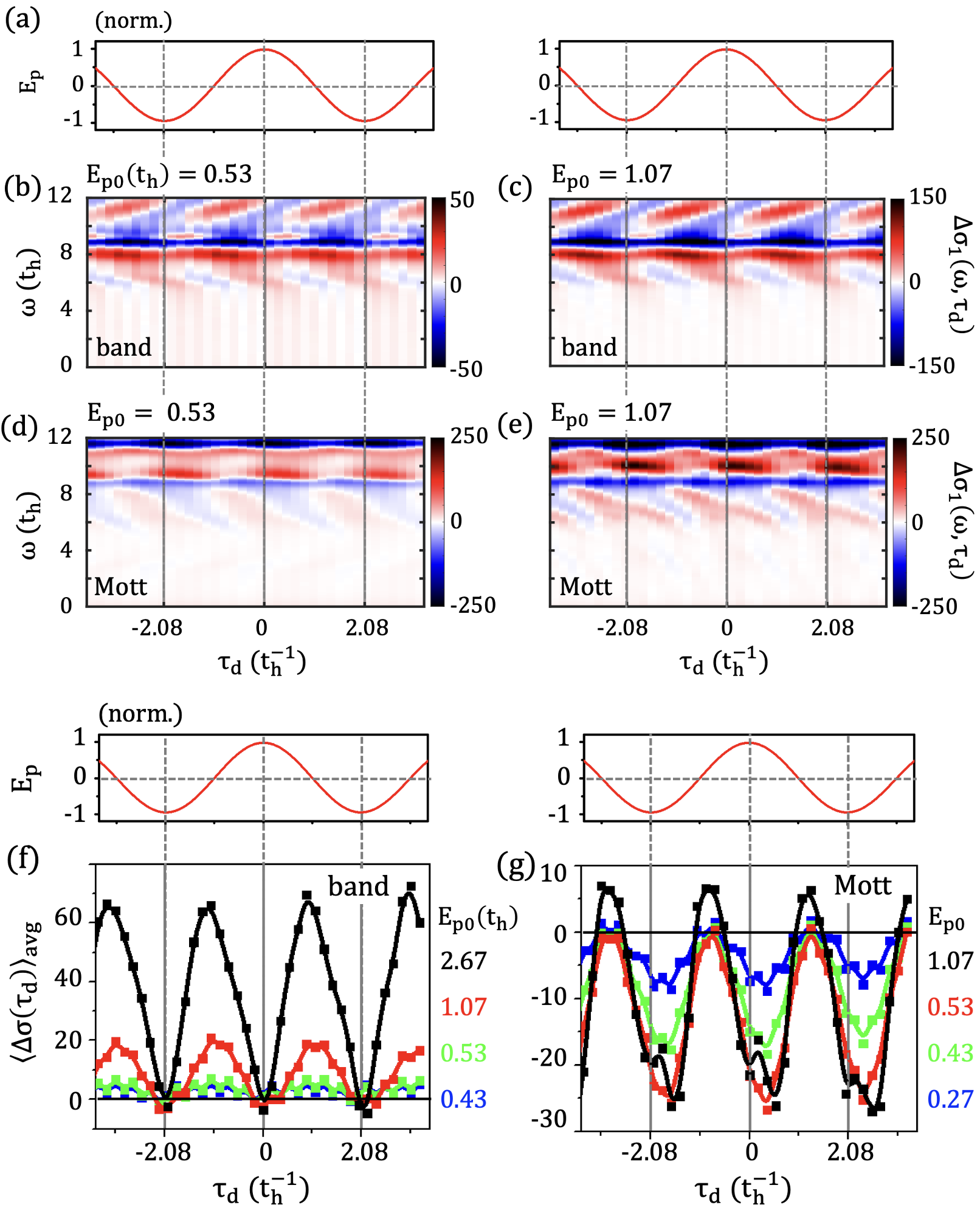}
\caption{
Calculated difference spectra of the pump-probe ATAS and the averaged ATAS for the band insulator and the Mott insulator. 
(a), Time-profile of the normalized electric field of pump pulse with frequency of $\omega_{p}/t_{h} = 1.5$, and $\omega_{p} \ll \Delta$ and $\omega_{p} \ll U$.
(b)-(e), $\Delta\sigma_{1}$ as a function of $\tau_{d}$ of the band insulator (b),(c) and the Mott insulator (d),(e) under different peak electric field strengths $E_{p0}$.
(f),(g), The averaged ATAS below the band edge $\langle \Delta \sigma(\tau_d)\rangle_{avg}$ for the band insulator (f) and the Mott insulator (g). 
The average is performed by $\langle \Delta \sigma(\tau_d)\rangle_{avg} = 1/\epsilon\int_{\epsilon_{E}-\epsilon}^{\epsilon_{E}} d\omega \Delta \sigma_{1}(\omega,\tau_d)$ at every given $\tau_d$. The range of the average is $\epsilon/t_{h} = 2$, and the band edge is $\epsilon_{E}/t_{h}=9$ for the band insulator (f) and the $\epsilon_{E}/t_{h}=8.8$ for the Mott insulator (g).
}
\label{FIG2}
\end{figure}

Next, we applied the ATAS to explorer a real time modification in the transient spectra shown in the Fig.\ref{FIG2}.
In the calculation, the vector potential contains both pump and probe field\cite{Sato1}, i.e., $A(\tau)=A_{p}(\tau)+A_{pr}(\tau-\tau_{d})$ with the pump-probe time-delay $\tau_{d}$. 
The first term describes the single packet of the pump pulse as $A_{p}(\tau)=A_{p0}\sin(\omega_{p}\tau)\cos^{2}(0.5\pi\tau/d)$ for $|\tau|\leq d$ and $A_{p}(\tau)=0$ for otherwise. The pump length parameter for an ultrashort pulse is fixed as $d = 9\pi/\omega_{p}$ to include $9$ periods of optical cycles in a single packet. 
The transient current becomes $J_{tr}(\tau,\tau_d)=J_{{p}+{pr}}(\tau,\tau_d)-J_{{p}}(\tau)$. The $J_{{p}+{pr}}(\tau,\tau_d)$ should be the current density when the ${\cal H}_{B(M)}(\tau)$ is evolved by the vector potential $A(\tau)=A_{p}(\tau)+A_{pr}(\tau-\tau_{d})$ and similarly, $J_{{p}}(\tau)$ under ${\cal H}_{B(M)}(\tau)$ of the $A(\tau)=A_{p}(\tau)$. 
Finally, the difference spectra of ATAS, which we shall call just ATAS hearafter, is practically obtained as\cite{Sato1},

\begin{eqnarray} 
\Delta \sigma_{1}(\omega,\tau_d) = Re[J_{tr}(\omega,\tau_d)/E_{pr}(\omega,\tau_d)]-\sigma_{1}(\omega)
\end{eqnarray}

and the probe electric field is $E_{pr}(\tau,\tau_d) = -(1/c) \partial / \partial \tau A_{pr}(\tau-\tau_{d})$.
In Fig.\ref{FIG2}(a), the time-profile of the pump electric field is shown with a relation $E_{p}(\tau_{d}) = -(1/c) \partial / \partial \tau A_{p}(\tau)|_{\tau=\tau_{d}}$.
In the Fig.\ref{FIG2}(b),(c), the ATAS from the band insulator exhibits two properties; (i) the redshift below the band edge and (ii) the fishbone structures, these results typically represent the fingerprint of DFKE\cite{Lucchini2,Sato1,Otobe1}. 
In the Fig.\ref{FIG2}(d),(e), for the Mott insulator, however, the distinctively nontrivial ATAS compared to the band insulator is captured.
It is interesting to note that while the fishbone-like structure still appears mimicking the DFKE, a blueshift bearing the negative difference absorption below the band edge is observed.
In addition, the responses of the phase oscillations are robust regardless of the field strength of the pump pulse.

In order to deeply understand the dynamics recorded by the ATAS from the Mott insulator in comparison with the band insulator, we show the averaged ATAS below the band edge in the Fig.\ref{FIG2}(f),(g).
In both systems, the results of the averaged ATAS commonly exhibit the $2\omega_p$-oscillations of its phases in the Mott insulator as does in the band insulator. Furthermore, the phase delays between the ATAS and the pump electric field are found to be out-of-phase, meaning the current responses are primarily governed by the dynamical nature\cite{Otobe1} (see the FIGS.1 and 2\cite{SM}).

On the other hand, in terms of the sign of spectra, the results clearly distinguish the band insulator and the Mott insulator.
The averaged ATAS from the band insulator completely shows a positive sign corresponding to the DFKE, since there are allowed probe channels assisted by the pump, as shown in the Fig.\ref{FIG2}(f). 
In contrast, in the Mott insulator in the Fig.\ref{FIG2}(g), it is remarkable that the averaged ATAS has a mostly negative sign, which does not obey the DFKE. 
Therefore, we would emphasize the unconventional ATAS found in the Mott insulator is observed to be nontrivial and thus, it is necessary to establish the underlying physics when the pure electron-correlation is incorporated into the ATAS.

When the electric field strength is more strongly enhanced, the phase delay of the band insulator gradually decreases and becomes in-phase, which is featured in the adiabatic regime $\gamma_{a} \gg 1$(see the FIGS.2\cite{SM}).
In the Mott insulator, however, the phase delay in this regime ($\gamma_{a} \gg 1$) cannot be simply defined. Since the structureless signals are introduced by the strong electron-correlations, the adiabatic responses of the FKE would not be monitored in the Mott insulator(see the FIGS.2 and 3\cite{SM}).

\begin{figure}
\includegraphics[width=0.47\textwidth]{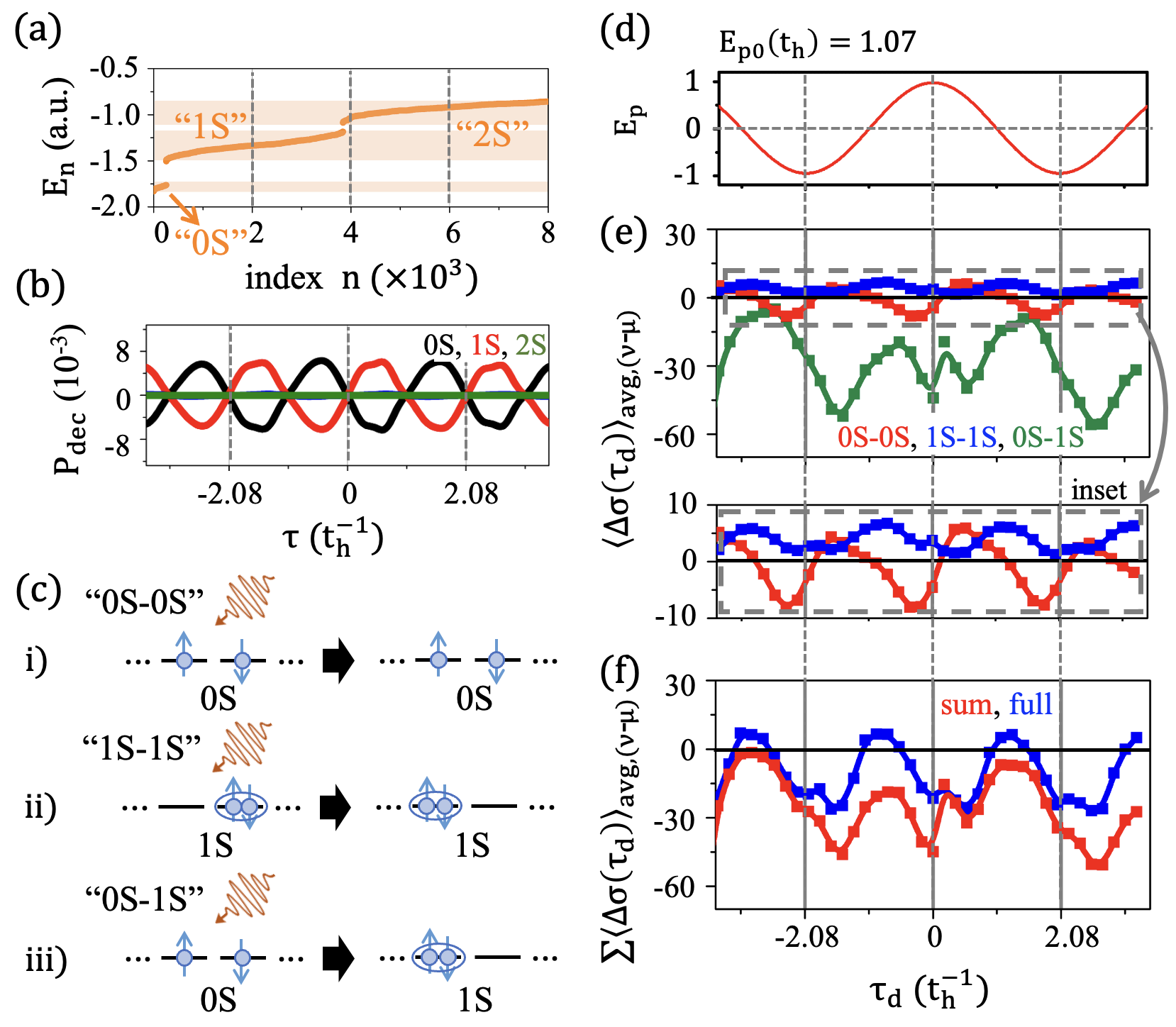}
\caption{
Decomposed ATAS for the Mott insulator with $E_{p0}(t_{h})=1.07$.
(a), Eigenvalues $E_{n}$ corresponding to each index number $n$ of the Mott insulator.
(b), The transition rate between decomposed set of eigenstates. 
(c), The schematics of main transition pathways for the decomposed ATAS, i.e., $0S$-$0S$, $1S$-$1S$, and $0S$-$1S$.
(d), The time-profile of the pump pulse.
(e), The averaged decomposed ATAS, $\langle \Delta \sigma(\tau_d)\rangle_{avg,(\nu\text{-}\mu)}$, of $\nu$-$\mu$ for $0S$-$0S$, $1S$-$1S$, and $0S$-$1S$. The inset is the zoomed results of the $0S$-$0S$, $1S$-$1S$.
(f), The summation of the averaged decomposed ATAS, $\sum \langle \Delta \sigma(\tau_d)\rangle_{avg,(\nu\text{-}\mu)}$, including components of $0S$-$0S$, $1S$-$1S$, and $0S$-$1S$. Note, the full calculation is from the Fig.\ref{FIG2}(g).
}
\label{FIG3}
\end{figure}

As seen from the Figs.\ref{FIG2}, the important feature separating the band insulator and the Mott insulator is the sign of the ATAS below the band edge. 
Thus we establish the relevant dynamical links between the sign of ATAS and the electron-correlation. 
We apply the decomposed calculation of the ATAS for the Mott insulator shown in the Fig.\ref{FIG3}. 
In the Fig.\ref{FIG3}(a), the eigenvalues of $n$th index, $E_{n}$, with corresponding eigenstates following the ${\cal H_M}|\Psi_{n}\rangle = E_{n}|\Psi_{n}\rangle$ are depicted. 
As the index $n$ increases, the eigenvalue also continuously increases. 
As seen in the previous study\cite{Lee1}, one can see some discontinuous points dividing the groups of eigenstates with respect to the number of double occupancies.
Therefore, the $0S$, $1S$, $2S$, and so on correspond to the groups containing the eigenstates with a given number of double occupancies; $0$, $1$, $2$, and so on, respectively.
To investigate the main transition pathways related to the number of double occupancies, we compute the transition rate between decomposed eigenstates, $P_{dec}(\tau) = \partial/\partial\tau \sum_{n\in \nu}|\langle\Psi_n|\Psi(\tau)\rangle|^{2}$ with $\nu$ indicating each group, i.e., $0S$, $1S$, and $2S$, as shown in the Fig.\ref{FIG3}(b). 
From the results, the significant transition pathways are observed to mostly include the $0S$ and $1S$. 
The pathways containing the $2S$ or higher terms should be rarely important in the present dynamics. 
Therefore, we will mainly focus on the transition pathways made of $0S$ and $1S$ as displayed in the Fig.\ref{FIG3}(c). 
The transition between the $0S$ and $1S$ should accompany the change in number of double occupancies, while the others conserve their numbers.
One can consequently decompose the ATAS into the three transition groups: first, transitions comprising the number-conserved double occupancy ($0S$-$0S$ and $1S$-$1S$), and second, transitions comprising the number-changed double occupancy ($0S$-$1S$), respectively.

From the above discussion, one can define the decomposed ATAS by following the new current operator $\bar{j}(\tau)_{(\nu\text{-}\mu)}$ on the basis of $j(\tau)_{\nu,\mu}$, we have
\begin{eqnarray} 
j(\tau) \rightarrow j(\tau)_{\nu,\mu}=\sum_{n\in\nu,m\in\mu}|\Psi_n\rangle\langle\Psi_n|j(\tau)|\Psi_m\rangle\langle\Psi_m|
\end{eqnarray} 
with $\nu$ and $\mu$ belonging to the groups; $0S$ and $1S$, that is, the number-conserved parts will be $\bar{j}(\tau)_{(0S\text{-}0S)}=j(\tau)_{0S,0S}$ and $\bar{j}(\tau)_{(1S\text{-}1S)}=j(\tau)_{1S,1S}$, and the number-changed part $\bar{j}(\tau)_{(0S\text{-}1S)}=j(\tau)_{1S,0S}+j(\tau)_{0S,1S}$.
From the results of $\bar{j}(\tau)_{(\nu\text{-}\mu)}$, the decomposed ATAS can be rewritten as follows, $\Delta\sigma_{1}(\omega,\tau_d) \rightarrow \Delta\sigma_{1}(\omega,\tau_d)_{(\nu\text{-}\mu)}$ and eventually, the averaged decomposed ATAS $\langle \Delta \sigma(\tau_d)\rangle_{avg,(\nu\text{-}\mu)}$ can be computed. 
In the Fig.\ref{FIG3}(e), we represent the $\langle \Delta \sigma(\tau_d)\rangle_{avg,(\nu\text{-}\mu)}$ of $\nu$-$\mu$ for $0S$-$0S$, $1S$-$1S$, and $0S$-$1S$. 
It is worth noting that the ATAS induced from the $0S$-$0S$ and $1S$-$1S$ are found to show partially positive and fully positive signs.
In contrast, the transitions for $0S$-$1S$ become special pathways since the strong electron-correlation brings about change in many-body energies $\sim U$ equivalent to forming a double occupancy.
It is discovered that the difference spectra of $0S$-$1S$ dominantly represent the negative sign, which entirely reproduces the full calculation (see the Fig.\ref{FIG3}(f)). 
As a result, we note that these pathways including $0S$-$1S$ mainly give rise to the negative sign to be the signature of the creation and the annihilation of double occupancy, clearly measured by ATAS.

\begin{figure}
\includegraphics[width=0.34\textwidth]{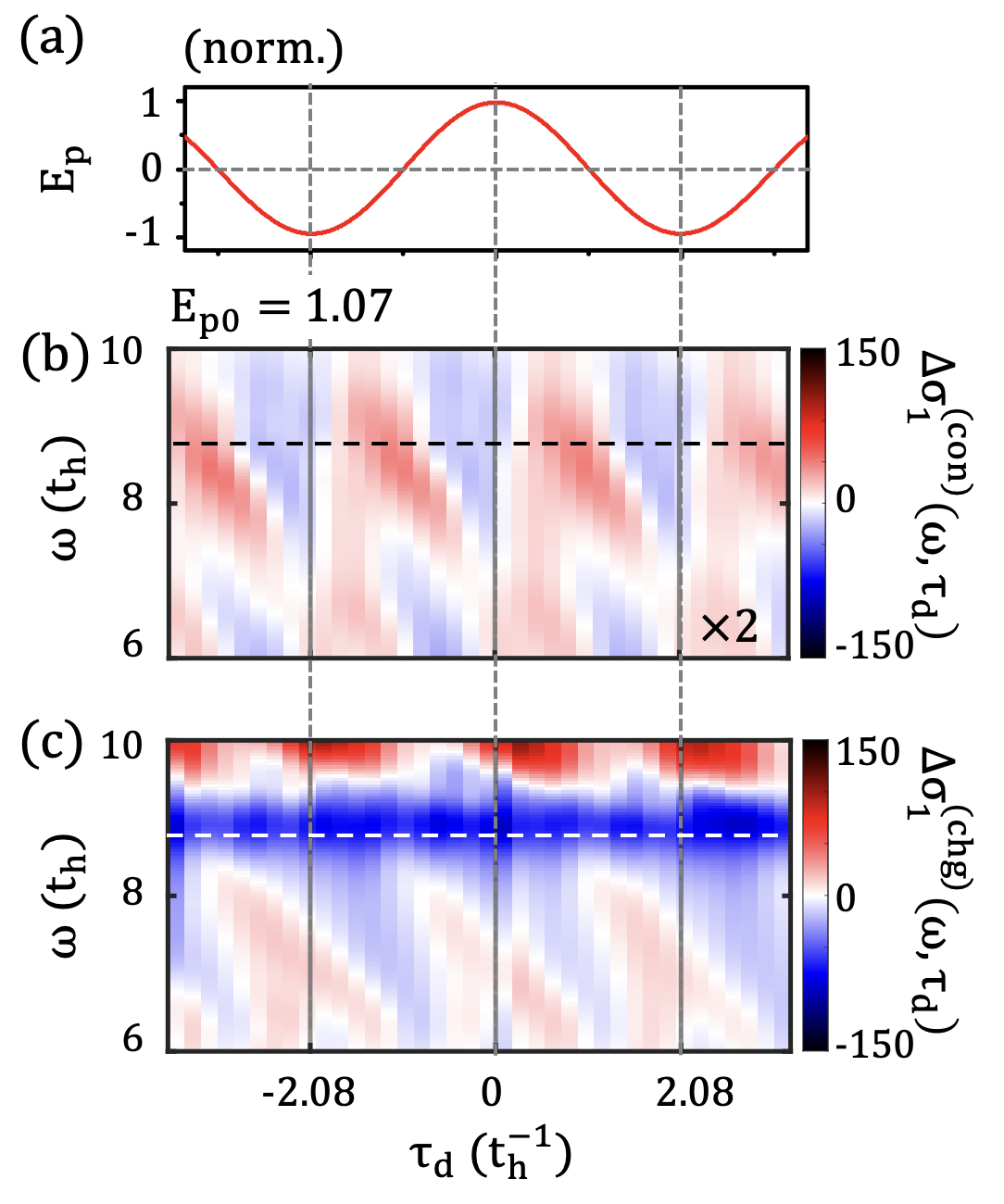}
\caption{
Decomposed ATAS of the Mott insulator in the $\omega\text{-}\tau_d$ with $E_{p0}(t_{h})=1.07$.
(a), Time-profile of the electric field of pump pulse.
(b), Decomposed ATAS containing the number-conserved transitions, $\Delta\sigma_{1}^{(con)}(\omega,\tau_d) = \Delta\sigma_{1}(\omega,\tau_d)_{(0S\text{-}0S)}+\Delta\sigma_{1}(\omega,\tau_d)_{(1S\text{-}1S)}$.
(c), Decomposed ATAS containing the number-changed transitions, $\Delta\sigma_{1}^{(chg)}(\omega,\tau_d) = \Delta\sigma_{1}(\omega,\tau_d)_{(0S\text{-}1S)}$.
}
\label{FIG4}
\end{figure}

Finally, to support our claims, we plot the decomposed ATAS in the $\omega\text{-}\tau_d$ plane as depicted in the Fig.\ref{FIG4}. 
In the Fig.\ref{FIG4}(b), the ATAS of number-conserved transitions $\Delta\sigma_{1}^{(con)}$ is found to have the positive signals below the band edge and the out-of-phase, which partially represents a feature of the DFKE as given in the Fig.\ref{FIG2}(b),(c). 
In the Fig.\ref{FIG4}(c), however, the ATAS of the number-changed transitions $\Delta\sigma_{1}^{(chg)}$ exhibits the highly negative signals below the band edge which quantitatively overwhelms the $\Delta\sigma_{1}^{(con)}$ due to the nontrivial excitations for strong electron-correlations\cite{Meinders,Kim} (see the FIGS.4\cite{SM}).
In addition, the silhouette of the Fig.\ref{FIG4}(c) also qualitatively reproduces the result of the full calculation.

In summary, we report the ATAS from band insulators and Mott insulators. 
In the band insulators, as commonly expected, the trivial responses are found accompanying the redshift below the band edge and the fishbone structures.
In the Mott insulator, however, where the system is associated with pure electron-correlation, the unconventional ATAS are observed to show the blueshift below band edge which violates the DFKE, but the fishbone-like structures still exist. 
It is also found that the transitions for the creation and the annihilation of double occupancy evidently contribute to the unconventional difference spectra leading to the blueshift.
Our study to establish the unconventional ATAS, incorporating the pure electron-correlation, guides us to a novel insight about correlation-embedded attosecond electron dynamics in condensed matters.

\section*{Acknowledgement}
We are grateful to J. D. Lee for the fruitful discussion.
The authors thank the computational support from the Center for Advanced Computation (CAC) at Korea Institute for Advanced Study (KIAS).
In this study, Y.K. supported by a KIAS Individual Grant (PG088601) at Korea Institute for Advanced Study (KIAS).
\\ \\
${}^{\ast}$ Corresponding author: ykim.email@gmail.com

\end{document}


\def\bb{\begin{equation}}
\def\ee{\end{equation}}

\title{Supplemental Material: Attosecond Transient Absorption Spectroscopy of Strongly Correlated Mott Insulators: Signature of the Creation and Annihilation of Double Occupancy}

\author{Youngjae Kim$^{*}$}

\affiliation{
School of Physics, KIAS, Seoul 02455, Korea
}

\date{\today}

\maketitle

\begin{figure}
\label{FIGS1}
\includegraphics[width=0.35\textwidth]{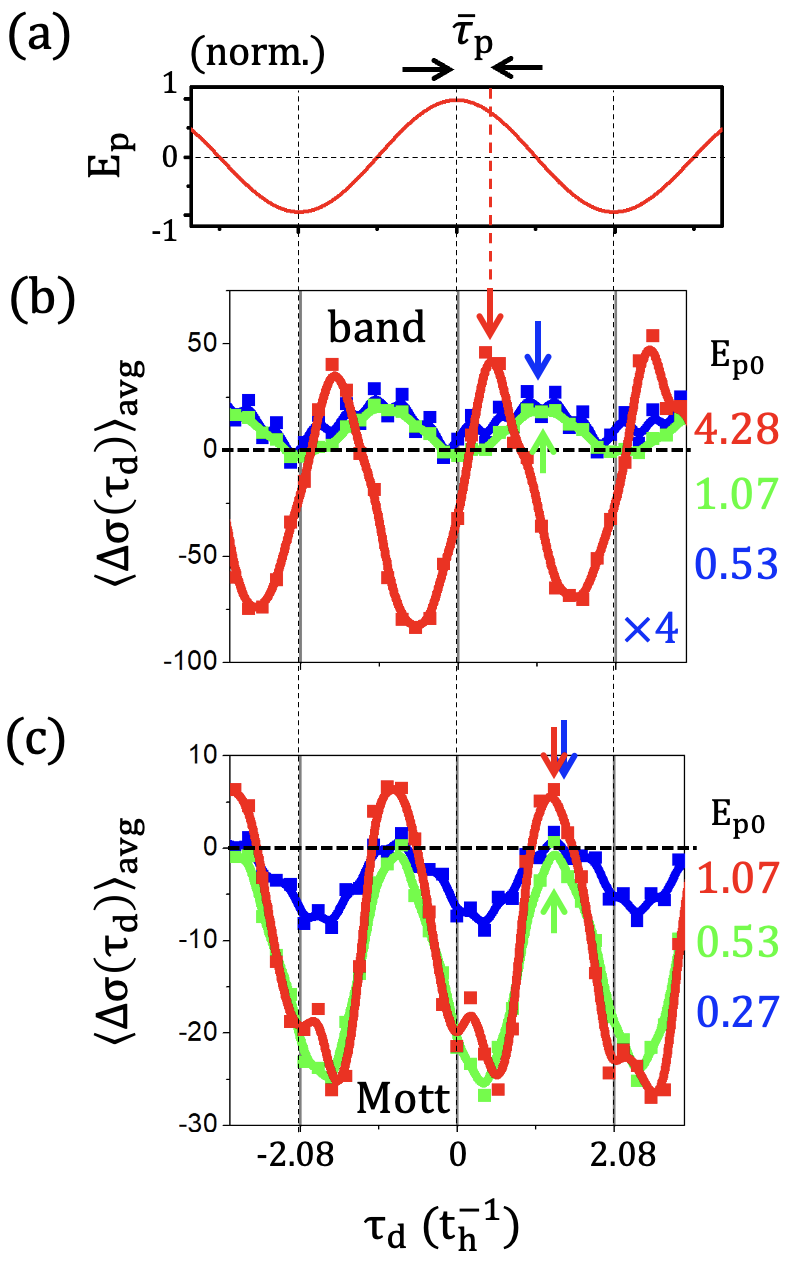}
\caption{
Calculated phase delay $\bar{\tau}_{p}$ between the averaged ATAS and the pump electric field.
(a), Time-profiles of the normalized electric field of pump pulse with frequency of $\omega_{p}/t_{h} = 1.5$ so that $\omega_{p} \ll \Delta$ and $\omega_{p} \ll U$.
(b), The averaged ATAS of the band insulator. 
(c), The averaged ATAS of the Mott insulator. 
In the (b),(c), the arrows indicate the peak of each averaged ATAS.
}
\end{figure}

In the FIGS.1, we begin to demonstrate the phase delay $\bar{\tau}_{p}$ between the averaged ATAS and the pump electric field for the band insulator and the Mott insulator. As seen from the FIGS.1(a), the averaged ATAS of the band insulator exhibits the peak shift depending on the pump field strength. As the pump field strength strongly increases, the peak is found to shift to the peak of electric field. In addition, the $2\omega_{p}$-oscillations are robust regardless to the field strength.

In the Mott insulator in the FIGS.1(b), however, the $2\omega_{p}$-oscillations are robust only below the $E_{p0} \sim 1$. In the regime of strong field, i.e., $E_{p0} \gg 1$, the ATAS of the Mott insulator does not show $2\omega_{p}$-oscillations (see the FIGS.3), which would originate from higher scattering natures such as higher double occupancies.

\begin{figure}
\includegraphics[width=0.3\textwidth]{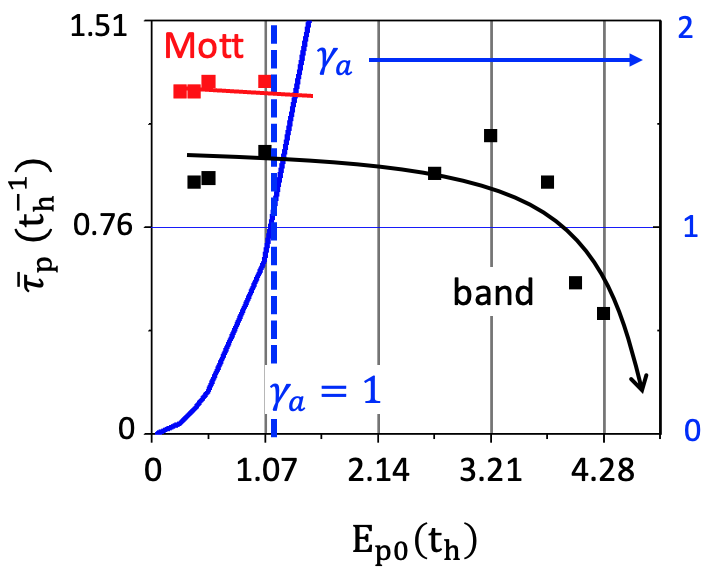}
\caption{
Estimated phase delay $\bar{\tau}_{p}$ as a function of pump field strength. 
The left (right) axis explains the $\bar{\tau}_p$ ($\gamma_{a}$). The dashed blue denotes the $\gamma_{a} = 1$.
}
\label{FIGS2}
\end{figure}

In the FIGS.2, we shall define the adiabatic parameter as $\gamma_{a} = U_{pon}/\omega_{p}$. 
The well-known ponderomotive energy will be $U_{pon}=A_{p0}^2/4\mu c^2$ and $A_{p0}$, $\mu$, and $\omega_{p}$ denote the peak of pump vector potential, reduced mass of the band insulator, and the pump frequency, respectively. Eventually, one can plot the phase diagram as depicted in the FIGS.2. 

For the band insulator, in the regime of $\gamma_{a} \sim 1$, the $\bar{\tau}_p$ remains $\sim 0.8 (t_h^{-1})$ corresponding to the out-of-phase between the ATAS and the pump electric field, belonging to the dynamical regime, i.e., the DFKE. When $\gamma_{a} \gg 1$, however, the $\bar{\tau}_p$ decreases to the $\sim 0 (t_h^{-1})$ corresponding to the in-phase of which the oscillations of ATAS are followed by the pump electric field. It indicates the dynamics belongs to the adiabatic regime, i.e., the FKE.

In case of the Mott insulator, the $\bar{\tau}_p$ can be defined only within the $\gamma_{a} \sim 1$ since when $\gamma_{a} \gg 1$ the higher scattering nature appears and overwhelms the $2\omega_{p}$-oscillations of ATAS. Therefore, we note that the result of the regime $\gamma_{a} \gg 1$ for the Mott insulator is too complicated to understand their dynamics. 
Thus it is out-of-scope in the present study.

\begin{figure}
\includegraphics[width=0.3\textwidth]{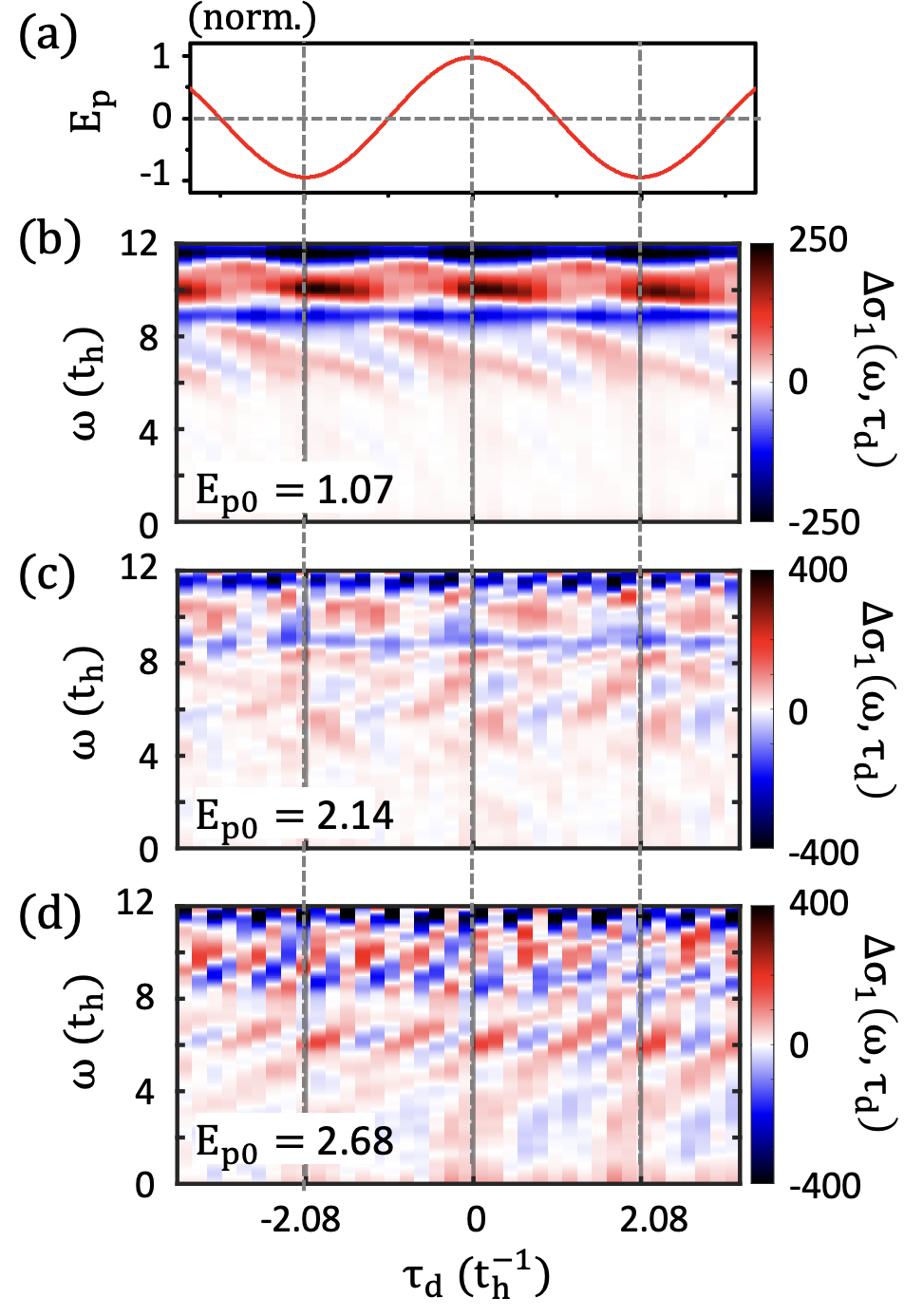}
\caption{
Calculated ATAS of the Mott insulator as a function of pump field strength.
(a), Time-profiles of the normalized electric field.
(b)-(d), $\Delta\sigma_{1}$ as a function of $\tau_{d}$ with various field strengths.
In the (d), under the strong pump field strength, the ATAS shows structureless signals beyond the $2\omega_{p}$-oscillations. 
}
\label{FIGS3}
\end{figure}

In the FIGS.3, we display the ATAS of the Mott insulator. In the regime of $\gamma_{a} \sim 1$ of the FIGS.3(b), the $2\omega_{p}$-oscillations of ATAS are robust so that the $\bar{\tau}_p$ can be defined as in the FIGS.1(c). In the regime of $\gamma_{a} \gg 1$ of the FIGS.3(d), the ATAS becomes structureless due to the higher scattering natures under strong laser field and thus, the $\bar{\tau}_p$ cannot be defined.

\begin{figure}
\includegraphics[width=0.4\textwidth]{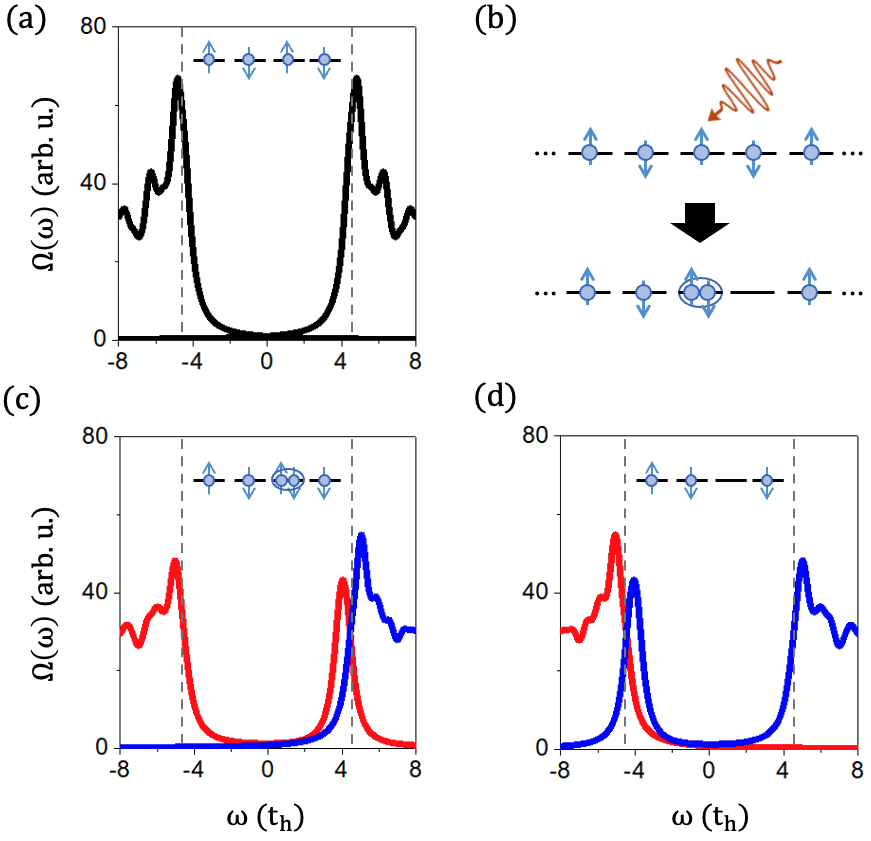}
\caption{
Calculated spectral weights of excited model of the Mott insulators.
(a), Spectral weight of the half-filled ($N$) groundstates.
(b), Schematics of the pump excitation and the creations of double occupancy.
(c),(d), Spectral weights of excited model for double occupancy.
The red and blue lines represent occupied and unoccupied states.
(c), Spectral weight of $N+1$ states for double occupancy.
(d), Spectral weight of $N-1$ states.
Creations and annihilations of double occupancy accompany change in the many-body energy from the Hubbard $U$ to modify spectral weights and optical properties.
}
\label{FIGS4}
\end{figure}

${}^{\ast}$ Corresponding author: ykim.email@gmail.com